# Visual observation of optical Floquet–Bloch oscillations

Zhen Zhang,[1,†] Yuan Li,[2,†] Xiankai Sun,[2,*] and Xuewen Shu[1,*]

[1]*Wuhan National Laboratory for Optoelectronics and School of Optical and Electronic Information, Huazhong University of Science and Technology, Wuhan, Hubei 430074, China*

[2]*Department of Electronic Engineering, The Chinese University of Hong Kong, Shatin, New Territories, Hong Kong SAR, China*

[†]*These authors contributed equally to this work*

[*]*Corresponding author: xksun@cuhk.edu.hk (X. Sun); xshu@hust.edu.cn (X. Shu)*

**Abstract:** Bloch oscillations, an important transport phenomenon, have extensively been studied in static systems but remain largely unexplored in Floquet systems. Here, we propose a new type of Bloch oscillations, namely the "Floquet–Bloch oscillations," which refer to rescaled Bloch oscillations with a period of extended least common multiple of the modulation and Bloch periods. We report the first visual observation of such Floquet–Bloch oscillations in femtosecond-laser-written waveguide arrays by using waveguide fluorescence microscopy. These Floquet–Bloch oscillations exhibit exotic properties, such as fractal spectrum and fractional Floquet tunneling. This new transport mechanism offers an intriguing method of wave manipulation, which has significant applications in coherent quantum transport.

**INTRODUCTION**

As a fundamental phenomenon of coherent quantum motion, Bloch oscillations, the oscillatory motion of a quantum particle with a BO period $\Lambda_{BO}$, were first predicted by Bloch and Zener in the context of crystal under a constant electric field (*1, 2*). Nevertheless, Bloch oscillations have never been experimentally observed in natural crystals owing to electron–phonon interactions. Compared with natural crystals, photonic systems exhibit strong interference effects, thus providing an alternative platform for investigating BOs. Optical Bloch oscillations, as an optical analog of Bloch oscillations, have been observed in diverse photonic systems such as coupled-waveguide arrays (*3-7*) and synthetic frequency lattices (*8-12*). Subsequently, abundant interesting phenomena concerning optical BOs have been theoretically predicted and experimentally observed in static systems, such as anharmonic Bloch oscillations (*13-15*), Bloch–Zener oscillations (*16-19*), and fractional Bloch oscillations (*20, 21*).

To date, the exploration of BOs in periodically driven systems (Floquet systems) has drawn tremendous attention because of their exotic characteristics (*22-38*). More specifically, two types of Bloch-like oscillations with harmonic modulation have been investigated as quasi-Bloch oscillations (QBOs) (*22-26*) and super-Bloch oscillations (SBOs) (*26-32*). QBOs refer to typical Bloch oscillations accompanied by sub-oscillations, where the BO period $\Lambda_{BO}$ is an integer multiple of the modulation period $\Lambda_{FL}$, i.e., $\Lambda_{BO} = N\Lambda_{FL}$ ($N$ is a positive integer greater than 1). SBOs refer to rescaled BOs with super large oscillation amplitude and period, where the BO period $\Lambda_{BO}$ (or its integer multiple) is slightly detuned from the modulation period $\Lambda_{FL}$, i.e., $\Lambda_{FL} \sim N\Lambda_{BO}$. While BOs in Floquet systems with several specific cases have been investigated, a general phenomenon concerning BOs in Floquet systems and the corresponding experimental observation remain largely elusive.

In this article, we explore the optical Bloch oscillations in Floquet systems and draw two essential conclusions: (1) Floquet lattice with a period $\Lambda_{FL}$ in a linear tilted potential leading to BOs with a period $\Lambda_{BO}$ can also be mapped onto another Floquet lattice with a period $\Lambda_{FBO}$ of the extended least common multiple (LCM) of $\Lambda_{FL}$ and $\Lambda_{BO}$; (2) When $\Lambda_{FL} \neq N\Lambda_{BO}$, the Floquet–Bloch oscillations occur with a FBO period of LCM($\Lambda_{FL}$, $\Lambda_{BO}$); when $\Lambda_{FL} = N\Lambda_{BO}$, spreading usually occurs. We emphasize that all the above conclusions are tenable for arbitrary Floquet engineering with a rational ratio of $\Lambda_{BO}/\Lambda_{FL}$. Therefore, Floquet–Bloch oscillations are a unified phenomenon of the existing Bloch oscillations, namely super-Bloch oscillations ($\Lambda_{FL} \sim N\Lambda_{BO}$) and quasi-Bloch oscillations ($\Lambda_{BO} = N\Lambda_{FL}$).

We experimentally verified our prediction in one-dimensional curved waveguide arrays fabricated with femtosecond laser writing technology. With waveguide fluorescence microscopy, we directly visualized the breathing and oscillatory motions of Floquet–Bloch oscillations. We provided a detailed analysis of FBOs, and investigated fractal spectrum and fractional Floquet tunneling. More specifically, we found that the FBO period $\Lambda_{FBO}$ is the Thomae's function (a fractal spectrum) of the ratio $\Lambda_{BO}/\Lambda_{FL}$, and several peaks of such a fractal spectrum were experimentally confirmed. In addition, the modulation-induced rescaling of the FBO amplitude depends largely on the ratio $\Lambda_{BO}/\Lambda_{FL}$, which refers to fractional Floquet tunneling. By varying the amplitude of harmonic modulation, we experimentally demonstrated that such rescaling of FBO amplitude follows a linear combination of fractional-order Anger and Weber functions. Our demonstration provides a promising method for controlling wave transport in photonics with potential applications in self-imaging, optical communication, and photonic quantum simulations.

**RESULTS**

**Theory of Bloch oscillations in a Floquet lattice**

Here, we employ a femtosecond-laser-written waveguide array in a fused silica substrate (Corning 7980) as an experimental platform for visualizing optical BOs in a Floquet systems (*39-41*). As depicted in Fig. 1A, we first considered a curved photonic lattice that consists of identical waveguides with waveguide spacing $d$. In the transverse direction $x$, the center of each waveguide core varies along the longitudinal direction $z$ by following a combined trajectory according to $x_0(z) = x_{BO}(z) + x_{FL}(z)$, where $x_{BO}(z)$ is the circular bending term with a bend radius $R$ ($R \gg z$) and $x_{FL}(z)$ $= M(z)$ is the periodic bending term with a modulation period $\Lambda_{FL}$ and modulation function $M(z)$ that satisfies $M(z) = M(z + \Lambda_{FL})$. In the case of paraxial propagation along the longitudinal direction $z$, the envelope $\psi(x, y, z)$ of the optical field guided in this photonic lattice at operating wavelength $\lambda$ is governed by the Schrödinger-type equation:

$$i\frac{\partial \psi}{\partial z} = -\frac{1}{2k_0}\nabla^2 \psi - \frac{k_0 \Delta n(x, y, z)}{n_0}\psi, \quad (1)$$

where $\nabla^2 = \partial_x^2 + \partial_y^2$ is the Laplacian operator in the transverse plane, $n_0 \sim 1.46$ is the refractive index of the substrate, $k_0 = 2\pi n_0/\lambda$ is the wave number, and $\Delta n(x, y, z) = n(x, y, z) - n_0$ is the femtosecond-laser-induced refractive-index increase ($\Delta n > 0$) that defines the entire photonic lattice. By considering a reference coordinate frame where the waveguides are straight in the $\tilde{z}$ direction, namely: $\tilde{x} = x + x_0(z)$, $\tilde{y} = y$, and $\tilde{z} = z$, the paraxial equation in the transformed coordinates can be expressed as

$$i\frac{\partial \tilde{\psi}}{\partial \tilde{z}} = -\frac{1}{2k_0}\tilde{\nabla}^2 \tilde{\psi} - \frac{k_0}{n_0}[\Delta n(\tilde{x}, \tilde{y}, \tilde{z}) + F(\tilde{z})\tilde{x}]\tilde{\psi} \quad (2)$$

with $\tilde{\psi} = \psi(\tilde{x}, \tilde{y}, \tilde{z})\exp\left\{-\frac{ik_0}{2\pi}\partial_{\tilde{z}}x_0(\tilde{z})\tilde{x} - \frac{ik_0}{4\pi}\int_0^{\tilde{z}}[\partial_{\tilde{z}}x_0(\tau)]^2 d\tau\right\}$ and $F(\tilde{z}) = -n_0\partial_{\tilde{z}}^2 x_0(\tilde{z})$. The additional term $F(\tilde{z})$ is determined by the combined trajectory and can be separated into two terms, i.e., $F(\tilde{z}) = F_{BO} + F_{FL}$, with $F_{BO} = n_0/R$ and $F_{FL} = -n_0\partial_{\tilde{z}}^2 M(\tilde{z})$. Equation (2) indicates that the propagation of low-power light in the proposed lattice is analogous to the temporal evolution of noninteracting electrons in a periodic potential subject to an electric field, where the spatial coordinate $\tilde{z}$ acts as time $t$, $F(\tilde{z})$ plays the role of the electric field force $F(t)$, and the refractive index profile $[\Delta n(\tilde{x}, \tilde{y}, \tilde{z}) + F(\tilde{z})\tilde{x}]$ is related to the sign-reversed driven potential $-V(t, \tilde{x})$. As sketched in Fig. 1B, the effective potential $-(\Delta n(\tilde{x}, \tilde{y}, \tilde{z}) + F_{FL}\tilde{x})$ refers to a Floquet lattice, and the linear potential gradient $F_{BO}$ gives rise to BOs with a period $\Lambda_{BO} = \lambda R/(n_0 d)$. As a result, our proposed scheme provides an experimental realization of optical Bloch oscillations in a Floquet lattice.

Figure 1C displays the cross-sectional microscope image of a fabricated sample. Each waveguide in our sample supports a well-confined fundamental mode, allowing the application of nearest-neighbor tight-binding approximation, so the propagation of guided light can be described by the following set of coupled equations:

$$i\frac{\partial a_m}{\partial z} = -c_0(a_{m-1} + a_{m+1}) - \frac{k_0 m F(z) d}{n_0} a_m \quad (3)$$

where $a_m$ is the amplitude of guided mode $|m\rangle$ in the $m^{th}$ waveguide and $c_0$ is the coupling constant between the nearest-neighbor waveguides. In the absence of force $F(z)$, i.e., for straight waveguide arrays, inserting a Bloch function $\left|\psi_{m,k_x}\right\rangle = \frac{1}{\sqrt{N}}\sum_m e^{ik_x dm}|m\rangle$ yields the single-band dispersion $\beta(k_x)$ = $2c_0\cos(k_x d)$ (blue line in Fig. 1E), where $\beta(k_x)$ denotes the quasienergy and $k_x$ denotes the transverse Bloch momentum. According to the generalized acceleration theory (29), the presence of force $F(z)$ leads to a shift of the transverse Bloch momentum $k_x(z) = k_x(0) + \frac{k_0}{n_0}\int_0^z F(\tau)d\tau$ and the Houston function $\left|\psi_{m,k_x}(z)\right\rangle = \exp\left\{-\frac{ik_0}{n_0}\int_0^z \beta[k_x(\tau)]d\tau\right\}\left|\psi_{m,k_x(z)}\right\rangle$ is the reconstructed solution (see the Supplementary Materials). When $P\Lambda_{BO} = Q\Lambda_{FL}$ ($Q$, $P$ are mutually prime integers), the extended least common multiple (LCM) of $\Lambda_{BO}$ and $\Lambda_{FL}$ is defined as LCM($\Lambda_{BO}$, $\Lambda_{FL}$) = $P\Lambda_{BO} = Q\Lambda_{FL}$, and $\beta[k_x(z)]$ is a z-periodic function with a period $\Lambda_{FBO}$ = LCM($\Lambda_{FL}$, $\Lambda_{BO}$) (see the Supplementary Materials). Consequently, the integral of $\beta[k_x(z)]$ can be expressed as a sum of a linear function and a periodic function, i.e., $\int_0^z \beta[k_x(\tau)]d\tau = \varepsilon(k_x)z + P(z)$ with $P(z) = P(z + \Lambda_{FBO})$. As a result, the entire lattice can be mapped onto another Floquet lattice, since the Houston function can be reduced to Floquet states as

$$\left|\psi_{m,k_x}(z)\right\rangle = \exp\left[-\frac{iz}{k_0}\varepsilon(k_x)\right]\left|u_{m,k_x(z)}\right\rangle, \quad (4)$$

where $\left|u_{m,k_x(z)}\right\rangle = \exp\left\{-\frac{i}{k_0}\int_0^z \beta[k_x(\tau)] - \varepsilon(k_x)d\tau\right\}\left|\psi_{m,k_x(z)}\right\rangle = \left|u_{m,k_x(z+\Lambda_{FBO})}\right\rangle$ is known as the Floquet function and $\varepsilon(k_x) \equiv \frac{1}{\Lambda_{FBO}}\int_0^{\Lambda_{FBO}} \beta[k_x(\tau)]d\tau$ is the corresponding Floquet dispersion that provides the effective transport properties over a period $\Lambda_{FBO}$. Under the single band approximation, the Floquet dispersion is expressed as

$$\varepsilon(k_x) = \sum_{n=1}^{\Lambda_{FBO}/\Lambda_{FL}} \cos\left(\frac{2\pi\Lambda_{FL}}{\Lambda_{BO}}n\right)D(k_x), \quad (5)$$

where $D(k_x) = \frac{1}{\Lambda_{\text{FBO}}} \int_0^{\Lambda_{\text{FL}}} \cos\left[k_x(0)d - \frac{2\pi d\tau}{\Lambda_{\text{BO}}} + \partial_\tau M(-\tau)d\right] d\tau$ in general contributes nonflat dispersion. Equation (5) implies that there are two possibilities for optical Bloch oscillations in a Floquet system. When $\Lambda_{\text{FL}} \neq N\Lambda_{\text{BO}}$, a complete cancellation of all orders of diffraction $\sum_{n=1}^{\Lambda_{\text{FBO}}/\Lambda_{\text{FL}}} \cos\left(\frac{2\pi\Lambda_{\text{FL}}}{\Lambda_{\text{BO}}}n\right) = 0$ results in flat Floquet dispersion $\varepsilon(k_x) \equiv 0$, indicating that the state experiences a periodic motion and returns to the initial state after propagating a period $\Lambda_{\text{FBO}}$. We call this phenomenon "Floquet–Bloch oscillations," because it is a combined phenomenon of Floquet engineering and Bloch oscillation. When $\Lambda_{\text{FL}} = N\Lambda_{\text{BO}}$, the Floquet dispersion $\varepsilon(k_x) \equiv \frac{1}{\Lambda_{\text{FL}}} \int_0^{\Lambda_{\text{FL}}} \cos\left[k_x(0)d - \frac{2\pi\tau d}{\Lambda_{\text{BO}}} + \partial_\tau M(-\tau)d\right] d\tau$ is in general no longer flat and the state experiences spreading. We emphasize that the above conclusions are valid for an arbitrary modulation function $M(z)$. In this connection, the existing BOs under harmonic modulation, namely QBOs ($\Lambda_{\text{BO}} = N\Lambda_{\text{FL}}$) and SBOs ($\Lambda_{\text{FL}} \sim N\Lambda_{\text{BO}}$), are the specific cases of FBOs.

**Visual observation of Bloch oscillations in a Floquet lattice**

To illustrate the similarity and difference between Floquet–Bloch oscillations and the existing Bloch oscillations, we employed a harmonic modulation $M(z) = A\cos(2\pi z/\Lambda_{\text{FL}})$ (see Fig. 1D) with modulation amplitude $A$. Without loss of generality, we considered four specific scenarios that correspond to typical BOs ($A = 0$), QBOs ($\Lambda_{\text{BO}}/\Lambda_{\text{FL}} = 3$), SBO-like oscillations ($\Lambda_{\text{BO}}/\Lambda_{\text{FL}} = 4/3$), and spreading ($\Lambda_{\text{BO}}/\Lambda_{\text{FL}} = 1$). The corresponding shifts of the transverse Bloch momentum $k_x(z) = k_x(0) + \frac{k_0}{n_0}\left[\frac{2\pi z}{\Lambda_{\text{BO}}} + \frac{2\pi A k_0}{\Lambda_{\text{FL}}}\sin\left(\frac{2\pi z}{\Lambda_{\text{FL}}}\right)\right]$ are displayed in Fig. 1F, where the harmonic modulation contributes a sub-oscillation to the states with Bloch-momentum-oscillation amplitude $(2\pi A k_0)/\Lambda_{\text{FL}}$. In the latter three scenarios, we considered the modulation amplitude $A = A_0\Lambda_{\text{FL}}/\Lambda_{\text{BO}}$ so that the sub-oscillation amplitude was normalized to $(2\pi A_0 k_0)/\Lambda_{\text{BO}}$. To experimentally verify our prediction, we fabricated a set of 90-mm-long samples composed of 31 identical waveguides with a waveguide spacing $d = 16$ μm. With such a waveguide spacing $d$, the coupling coefficient between straight waveguides $c_0 \sim 1.45$ cm$^{-1}$ was experimentally characterized. These waveguides follow the combined trajectories having a bend radius $R = 110.8$ cm (corresponding to $\Lambda_{BO} \sim 30$ mm) and the modulation period $\Lambda_{FL} = 10$, 22.5, and 30 mm (corresponding to the ratios $\Lambda_{BO}/\Lambda_{FL} = 3$, 4/3, and 1, respectively). With the considered modulation period, $A_0 = 18$ μm was chosen to reduce the associated radiation losses of waveguides. Further details of the fabrication processes are provided in the Supplementary Materials.

Similar to the existing Bloch oscillations, Floquet–Bloch oscillations exhibit a breathing and an oscillatory motion under a single-site excitation and a broad-beam excitation, respectively. In the following experiments, we implemented visible-light excitation ($\lambda = 633$ nm) and directly visualized both the breathing modes and oscillating modes of Floquet–Bloch oscillations by using waveguide fluorescence microscopy (see the Supplementary Materials). A coordinate transformation that maps circular arcs into straight lines was applied to digitally process the fluorescence image so that the light evolution could be visualized more intuitively.

First, we focus on the breathing modes under a single-site excitation. The narrow excitation in the real space corresponds to a broad excitation of Bloch modes in the reciprocal space, resulting in strongly diffracting wave packets. To quantify the diffraction of wave packets for the single-site excitation, we define the variance of excitation at the distance $z$ in such a discrete system as

$$\sigma^2(z) = \frac{\sum_m m^2 |a_m|^2}{\sum_m |a_m|^2}. \tag{5}$$

The light is initially excited in the central waveguide resulting in a vanishing variance $\sigma^2(0) = 0$, and a rise of the variance indicates that the light experiences broadening. Under the single-site excitation, the experimental results, respective simulations, and extracted variances $\sigma^2(z)$ for the scenarios considered in Fig. 1F are summarized in Fig. 2, where Figs. 2 (A to C), (D to F), (G to I), (J to H) corresponding to typical BOs, QBOs, SBO-like oscillations, and spreading, respectively. Without modulation ($A = 0$), Figs. 2 (A and B) displays the light evolution that corresponds to typical BOs, where the measured BO period ~30 mm is consistent with its theoretical value $\Lambda_{BO} = R\lambda/(n_0 d)$. The light first broadens until it propagates half of the BO period and then focuses into the central waveguide again at the BO period, as $\sigma^2$ reaches its maximum at $z \sim 15$ mm and then decreases to zero at $z \sim 30$ mm (see Fig. 2C). When the modulation is introduced, Bloch oscillations in the Floquet lattice exhibit diverse transport properties as expected, where the ratio $\Lambda_{BO}/\Lambda_{FL}$ makes a significant difference. For $\Lambda_{BO}/\Lambda_{FL} = 3$, the FBOs are observed and reduce into QBOs, where the FBO period $\Lambda_{FBO}$ is equal to the BO period $\Lambda_{BO}$ (see Figs. 2D and 2E). The QBOs pattern is basically similar to that of typical BOs, except that light experiences additional sub-oscillations, as $\sigma^2$ oscillates with dual periods (see Fig. 2F). For $\Lambda_{BO}/\Lambda_{FL} = 4/3$, the FBOs exhibit their similarity to SBOs, where the FBO period $\Lambda_{FBO} \sim 90$ mm is much longer than the BO period $\Lambda_{BO}$ (see Figs. 2G and 2H). In addition to the extended FBO period, we also observed dramatic broadening of the light, as the maximum of $\sigma^2$ is far larger than that of typical BOs (see Fig. 2I). For $\Lambda_{BO}/\Lambda_{FL} = 1$, the evolution of light propagating from 0 to $\Lambda_{FBO}/2$ cannot be cancelled with that propagating from $\Lambda_{FBO}/2$ to $\Lambda_{FBO}$. As a result, the condition for FBOs is destroyed and spreading occurs, where light exhibits ballistic spreading and is no longer localized (see Figs. 2J and 2K). The typical discrete diffraction pattern accompanied by oscillations is observed, as $\sigma^2$ oscillates around the gray-dashed curve of $2c_0^2 B_1^2 \left(\frac{2\pi A d k_0}{\Lambda_{FL}}\right) z^2$ where $B_1$ is the first-order Bessel function (see Fig. 2L).

Next, we focus on the oscillation modes under a broad-beam excitation. The broad-beam excitation in the real space corresponds to a narrow excitation in the reciprocal space. In this case, the group velocity of beam motion in the lattices can be expressed as $V_{group}(z) = -d\beta(z)/dk_x(z) = 2dc_0\sin(k_x(z)d)$, and the transverse displacement $\Delta x(z)$ of beam center is determined by $\Delta x(z) = \int_0^z V_{group}(\tau)d\tau$. Here we define the weighted-average position of excitation at the distance $z$ in such a discrete system as

$$x(z) = \frac{\sum_m md|a_m|^2}{\sum_m |a_m|^2}. \tag{6}$$

The excitation is located at the center of the lattice, i.e., $x(0) = 0$. During propagation, a rise (drop) of $x(z)$ indicates that the light shifts toward the $x$ ($-x$) direction. Here, we launched a 7-waveguide-wide Gaussian beam at normal incidence to the edge of the substrate. This corresponds to a narrow spectrum centered at $k_x(0) = 0$ in the reciprocal space. Under the broad excitation, the experimental results, respective simulations, and extracted trajectories of the beam $x(z)$ (white dashed lines) for the scenarios considered in Fig. 1F are summarized in Fig. 3, where Fig. 3 (A and B), (C and D), (E and F), and (G and H) correspond to BOs, QBOs, SBOs-like oscillations, and spreading, respectively. Without modulation ($A = 0$), Fig. 3 (A and B) display the light evolution that corresponds to typical BOs, where the broad beam undergoes a sinusoidal oscillation with the BO period $\Lambda_{BO}$. Similar to the breathing motion discussed previously, the oscillating motion exhibits diverse transport properties when the modulation is introduced. For $\Lambda_{BO}/\Lambda_{FL} = 3$, Fig. 3 (C and D) display the light evolution that corresponds to QBOs, where the trajectory of the broad beam follows a doubly oscillating function. The broad beam evolves along the $x$ direction and returns to the initial position after propagating any multiple of the BO period $\Lambda_{BO} \sim 30$ mm. For $\Lambda_{BO}/\Lambda_{FL} = 4/3$, Fig. 3 (E and F) display the light evolution that corresponds to SBOs-like oscillations, where the trajectory of the broad beam follows a giant doubly oscillating function with an extended period of ~90 mm. The maximal displacement of the broad beam for SBOs-like oscillations is observed at half of the FBO period, i.e., $z \sim 45$ mm. For $\Lambda_{BO}/\Lambda_{FL} = 1$, Fig. 3 (G and H) display the light evolution that corresponds to spreading. Although the trajectory of the broad beam follows an oscillating function, beam broadening is observed during propagation. As a result, the beam does not return to the initial state of excitation and the condition for FBOs is destroyed.

For both single-site and broad-beam excitations, the visual observations of fluorescence images and quantitative analyses have excellent agreement with the respective simulation results. Therefore, our waveguide arrays are capable of accurately revealing BOs in Floquet lattices.

**Fractal spectrum and fractional Floquet tunneling**

We also made a quantitative analysis of FBOs. Firstly, we studied the dependence of $\Lambda_{BO}/\Lambda_{FBO}$ on $\Lambda_{BO}/\Lambda_{FL}$. As shown in Fig. 4A, the theoretically predicted FBO period $\Lambda_{FBO} = \text{LCM}(\Lambda_{BO}, \Lambda_{FL})$ determines that the FBO period spectrum follows the Thomae's function. One may find that the Thomae's function is a fractal structure composed of infinite discrete peaks, where the patterns exhibit self-similarity at increasingly smaller scales. Owing to limited sample lengths, we fabricated a set of samples with $\Lambda_{BO}/\Lambda_{FBO} \geq 1/6$, fixed $\Lambda_{BO} = 30$ mm, and varied $\Lambda_{FL}$ from 15 to 30 mm. As expected, we experimentally verified several peaks of such a fractal spectrum by fitting the measured and simulated variance $\sigma^2(z)$ under single-site excitation (see the Supplementary Materials). This fractal spectrum demonstrates the relationship between FBOs and SBOs. When $\Lambda_{BO}/\Lambda_{FL}$ approaches 1, the Thomae's function can be approximated to a continuous linear function, implying that the FBOs reduce into SBOs with a period $\Lambda_{FBO} = \Lambda_{FL}\Lambda_{BO}/(\Lambda_{BO} - \Lambda_{FL})$. Moreover, the existence of FBOs is confirmed even for a large $\Lambda_{BO}/\Lambda_{FL}$, which goes far beyond SBOs. In agreement with theoretical prediction, such a spectrum reveals the fractal nature of the FBOs.

Secondly, under single-site excitations, we defined the FBO amplitude as $\sigma^2(\Lambda_{FBO}/2)$ and studied the dependence of FBO amplitude on modulation amplitude $A$. We found that the harmonic modulation leads to a rescaling of FBO amplitude following the square of a linear combination of the Anger function $J_\nu\left(\frac{2\pi A d k_0}{\Lambda_{FL}}\right)$ and the Weber function $E_\nu\left(\frac{2\pi A d k_0}{\Lambda_{FL}}\right)$ with a fractional order $\nu = \Lambda_{FL}/\Lambda_{BO}$, which refers to fractional Floquet tunneling (see the Supplementary Materials). Note that the rescaling of FBO amplitude depends largely on $\Lambda_{BO}/\Lambda_{FL}$, which provides a flexible way to manipulate the light. Figure 4B displays two examples of such Floquet tunneling, including QBOs (red line, $\Lambda_{BO}/\Lambda_{FL} = 3$) and SBOs-like oscillations (blue line, $\Lambda_{BO}/\Lambda_{FL} = 4/3$). Each curve is normalized to unity at its maximum. For the QBOs, the theoretically predicted FBO amplitude has a characteristic $\left[2\cos(\pi/3)E_{1/3}\left(-\frac{2\pi A d k_0}{\Lambda_{FL}}\right) + 2\sin(\pi/3)J_{1/3}\left(-\frac{2\pi A d k_0}{\Lambda_{FL}}\right)\right]^2$ dependence on $A/\Lambda_{FL}$. By contrast, the Floquet tunneling for the SBOs-like oscillations exhibits a different behavior, where the FBO amplitude has a characteristic $8J_{3/4}\left(\frac{2\pi A d k_0}{\Lambda_{FL}}\right)^2$ dependence on $A/\Lambda_{FL}$. To verify our prediction, we fabricated two sets of samples with a varied modulation amplitude $A$ and extracted the corresponding variance $\sigma^2(z)$ from the measured fluorescence images. As expected, one finds that the measured FBO amplitude has excellent agreement with its theoretical prediction. For the QBOs, with increasing amplitude $A$ the FBO amplitude decreases before it reaches zero, indicating that the introduction of harmonic modulation will not broaden the FBO amplitude compared with the typical BOs ($A = 0$). For the SBOs-like oscillations, with increasing amplitude $A$ the FBO amplitude first increases to its maximum around $A = 22.5$ μm and then decreases. The detailed experimental results are provided in the Supplementary Materials.

**DISCUSSION**

In summary, we report the first visual observation of optical BOs in a Floquet lattice and the investigation of Floquet–Bloch oscillations. In addition to the above-discussed cases with a harmonic modulation, we emphasize that Floquet–Bloch oscillations occur for arbitrary Floquet engineering $M(z)$ and the corresponding experimental results are provided in the Supplementary Materials. The visual observation of Floquet–Bloch oscillations is a key to understanding the underlying transport mechanism, which has a significant impact on both fundamental research and practical applications. For fundamental research, our theoretical and experimental work enables the exploration of a branch of fundamental phenomena involving FBOs, such as the interplay between FBOs and binary lattices (*16*), non-Hermitian lattices (*42*), and optical nonlinearity (*43*). For practical applications, the demonstrated manipulation of optical waves can be implemented in synthetic dimensions of time (*44*), frequency (*45*), and angular momenta (*46, 47*), leading to applications in high-efficiency frequency conversion and signal processing.

# REFERENCES AND NOTES

1. F. Bloch, Über die Quantenmechanik der Elektronen in Kristallgittern. *Z. Phys.* **52**, 555–600 (1929).
2. C. Zener, A theory of the electrical breakdown of solid dielectrics. *Proc. R. Soc. A* **145**, 523–529 (1934).
3. U. Peschel, T. Pertsch, F. Lederer, Optical Bloch oscillations in waveguide arrays. *Opt. Lett.* **23**, 1701–1703 (1998).
4. G. Lenz, I. Talanina, C. M. de Sterke, Bloch oscillations in an array of curved optical waveguides. *Phys. Rev. Lett.* **83**, 963–966 (1999).
5. A. Block, C. Etrich, T. Limboeck, F. Bleckmann, E. Soergel, C. Rockstuhl, S. Linden, Bloch oscillations in plasmonic waveguide arrays. *Nat. Commun.* **5**, 3843 (2014).
6. M. Lebugle, M. Graefe, R. Heilmann, A. Perez-Leija, S. Nolte, A. Szameit, Experimental observation of N00N state Bloch oscillations. *Nat. Commun.* **6**, 8273 (2015).
7. Y. L. Xu, W. S. Fegadolli, L. Gan, M. H. Lu, X. P. Liu, Z. Y. Li, A. Scherer, Y. F. Chen, Experimental realization of Bloch oscillations in a parity-time synthetic silicon photonic lattice. *Nat. Commun.* **7**, 11319 (2016).
8. C. Bersch, G. Onishchukov, U. Peschel, Experimental observation of spectral Bloch oscillations. *Opt. Lett.* **34**, 2372–2374 (2009).
9. L. Yuan, S. Fan, Bloch oscillation and unidirectional translation of frequency in a dynamically modulated ring resonator. *Optica* **3**, 1014–1018 (2016).
10. H. Chalabi, S. Barik, S. Mittal, T. E. Murphy, M. Hafezi, E. Waks, Guiding and confining of light in a two-dimensional synthetic space using electric fields. *Optica* **7**, 506–513 (2020).
11. H. Chen, N. Yang, C. Qin, W. Li, B. Wang, T. Han, C. Zhang, W. Liu, K. Wang, H. Long, X. Zhang, P. Lu, Real-time observation of frequency Bloch oscillations with fibre loop modulation. *Light Sci. Appl.* **10**, 48 (2021).
12. G. Z. Li, Y. L. Zheng, A. Dutt, D. Y. Yu, Q. R. Shan, S. J. Liu, L. Q. Yuan, S. H. Fan, X. F. Chen, Dynamic band structure measurement in the synthetic space. *Sci. Adv.* **7**, eabe4335 (2021).
13. G. Wang, J. P. Huang, K. W. Yu, Nontrivial Bloch oscillations in waveguide arrays with second-order coupling. *Opt. Lett.* **35**, 1908–1910 (2010).
14. F. Dreisow, G. Wang, M. Heinrich, R. Keil, A. Tünnermann, S. Nolte, A. Szameit, Observation of anharmonic Bloch oscillations. *Opt. Lett.* **36**, 3963–3965 (2011).
15. J. Stockhofe, P. Schmelcher, Bloch dynamics in lattices with long-range hopping. *Phys. Rev. A* **91**, 023606 (2015).
16. F. Dreisow, A. Szameit, M. Heinrich, T. Pertsch, S. Nolte, A. Tunnermann, S. Longhi, Bloch-Zener oscillations in binary superlattices. *Phys. Rev. Lett.* **102**, 076802 (2009).
17. S. Ding, G. P. Wang, Nonreciprocal optical Bloch-Zener oscillations in ternary parity-time-symmetric waveguide lattices. *Appl. Phys. Lett.* **100**, 151913 (2012).
18. Y. Sun, D. Leykam, S. Nenni, D. Song, H. Chen, Y. D. Chong, Z. Chen, Observation of valley Landau-Zener-Bloch oscillations and pseudospin imbalance in photonic graphene. *Phys. Rev. Lett.* **121**, 033904 (2018).

19. Y. J. Chang, Y. H. Lu, Y. Y. Yang, Y. Wang, W. H. Zhou, X. W. Wang, X. M. Jin, Inhibition and reconstruction of Zener tunneling in photonic honeycomb lattices. *Adv. Mater.* **34**, 2110044 (2022).
20. S. Longhi, Photonic Bloch oscillations of correlated particles. *Opt. Lett.* **36**, 3248–3250 (2011).
21. G. Corrielli, A. Crespi, G. Della Valle, S. Longhi, R. Osellame, Fractional Bloch oscillations in photonic lattices. *Nat. Commun.* **4**, 1555 (2013).
22. J. Wan, C. M. de Sterke, M. M. Dignam, Dynamic localization and quasi-Bloch oscillations in general periodic ac-dc electric fields. *Phys. Rev. B* **70**, 125311 (2004).
23. S. Longhi, Bloch dynamics of light waves in helical optical waveguide arrays. *Phys. Rev. B* **76**, 195119 (2007).
24. A. Joushaghani, R. Iyer, J. K. S. Poon, J. S. Aitchison, C. M. de Sterke, J. Wan, M. M. Dignam, Quasi-Bloch oscillations in curved coupled optical waveguides. *Phys. Rev. Lett.* **103**, 143903 (2009).
25. W. Zhang, X. Zhang, Y. V. Kartashov, X. Chen, F. Ye, Bloch oscillations in arrays of helical waveguides. *Phys. Rev. A* **97**, 063845 (2018).
26. A. R. C. Buarque, M. L. Lyra, W. S. Dias, Bloch-like superoscillations and unidirectional motion of phase-driven quantum walkers. *Phys. Rev. A* **103**, 012222 (2021).
27. E. Haller, R. Hart, M. J. Mark, J. G. Danzl, L. Reichsoellner, H.-C. Naegerl, Inducing transport in a dissipation-free lattice with super Bloch oscillations. *Phys. Rev. Lett.* **104**, 200403 (2010).
28. K. Kudo, T. S. Monteiro, Theoretical analysis of super–Bloch oscillations. *Phys. Rev. A* **83**, 053627 (2011).
29. S. Arlinghaus, M. Holthaus, Generalized acceleration theorem for spatiotemporal Bloch waves. *Phys. Rev. B* **84**, 054301 (2011).
30. S. Longhi, G. Della Valle, Correlated super-Bloch oscillations. *Phys. Rev. B* **86**, 075143 (2012).
31. Z. Turker, C. Yuce, Super Bloch oscillation in a PT symmetric system. *Phys. Lett. A* **380**, 2260–2264 (2016).
32. C. Qin, L. Yuan, B. Wang, S. Fan, P. Lu, Effective electric-field force for a photon in a synthetic frequency lattice created in a waveguide modulator. *Phys. Rev. A* **97**, 063838 (2018).
33. C. E. Creffield, F. Sols, Directed transport in driven optical lattices by gauge generation. *Phys. Rev. A* **84**, 023630 (2011).
34. C. Sias, H. Lignier, Y. P. Singh, A. Zenesini, D. Ciampini, O. Morsch, E. Arimondo, Observation of photon-assisted tunneling in optical lattices. *Phys. Rev. Lett.* **100**, 040404 (2008).
35. S. Mukherjee, A. Spracklen, D. Choudhury, N. Goldman, P. Ohberg, E. Andersson, R. R. Thomson, Modulation-assisted tunneling in laser-fabricated photonic Wannier–Stark ladders. *New J. Phys.* **17**, 115002 (2015).
36. A. Joushaghani, R. Iyer, J. K. S. Poon, J. S. Aitchison, C. M. de Sterke, J. Wan, M. M. Dignam, Generalized exact dynamic localization in curved coupled optical waveguide arrays. *Phys. Rev. Lett.* **109**, 103901 (2012).
37. Y. Mizumoto, Y. Kayanuma, Double Landau-Zener interferometry in a dc-ac–driven Bloch-Zener oscillation. *Phys. Rev. A* **88**, 023611 (2013).
38. L. K. Upreti, C. Evain, S. Randoux, P. Suret, A. Amo, P. Delplace, Topological swing of Bloch oscillations in quantum walks. *Phys. Rev. Lett.* **125**, 186804 (2020).
39. R. R. Gattass, E. Mazur, Femtosecond laser micromachining in transparent materials. *Nat. Photonics* **2**, 219–225 (2008).

40. A. Szameit, S. Nolte, Discrete optics in femtosecond-laser-written photonic structures. *J. Phys. B-At. Mol. Opt. Phys.* **43**, 163001 (2010).
41. D. Z. Tan, Z. Wang, B. B. Xu, J. R. Qiu, Photonic circuits written by femtosecond laser in glass: improved fabrication and recent progress in photonic devices. *Adv. Photonics* **3**, 024002 (2021).
42. S. Weimann, M. Kremer, Y. Plotnik, Y. Lumer, S. Nolte, K. G. Makris, M. Segev, M. C. Rechtsman, A. Szameit, Topologically protected bound states in photonic parity–time-symmetric crystals. *Nat. Mater.* **16**, 433–438 (2017).
43. L. J. Maczewsky, M. Heinrich, M. Kremer, S. K. Ivanov, M. Ehrhardt, F. Martinez, Y. V. Kartashov, V. V. Konotop, L. Torner, D. Bauer, A. Szameit, Nonlinearity-induced photonic topological insulator. *Science* **370**, 701–704 (2020).
44. A. Regensburger, C. Bersch, B. Hinrichs, G. Onishchukov, A. Schreiber, C. Silberhorn, U. Peschel, Photon propagation in a discrete fiber network: an interplay of coherence and losses. *Phys. Rev. Lett.* **107**, 233902 (2011).
45. C. Z. Qin, F. Zhou, Y. G. Peng, D. Sounas, X. F. Zhu, B. Wang, J. J. Dong, X. L. Zhang, A. Alu, P. X. Lu, Spectrum control through discrete frequency diffraction in the presence of photonic gauge potentials. *Phys. Rev. Lett.* **120**, 133901 (2018).
46. J. Floss, A. Kamalov, I. S. Averbukh, P. H. Bucksbaum, Observation of Bloch oscillations in molecular rotation. *Phys. Rev. Lett.* **115**, 203002 (2015).
47. X. W. Luo, X. X. Zhou, J. S. Xu, C. F. Li, G. C. Guo, C. W. Zhang, Z. W. Zhou, Synthetic-lattice enabled all-optical devices based on orbital angular momentum of light. *Nat. Commun.* **8**, 16097 (2017).

**Acknowledgments:** The authors acknowledge valuable comments and suggestions from Dr. Jingwen Ma. **Funding:** This work was supported by National Key Research and Development Program of China (2018YFE0117400), National Natural Science Foundation of China (62275093), Research Grants Council of Hong Kong (No. 14209519, C4050-21E), and Group Research Scheme and Strategic Partnership Award for Research Collaboration offered by The Chinese University of Hong Kong. **Author contributions:** Y.L. developed the theory of Floquet–Bloch oscillations and performed the simulation. Z.Z. fabricated the device and performed the measurement. Y.L. and Z.Z. analyzed the data and wrote the paper. All authors discussed the results and commented on the manuscript. X.S. (Xuewen Shu) and X.S. (Xiankai Sun) supervised the project. **Competing interests:** The authors declare that they have no competing interests. **Data and materials availability:** All data needed to evaluate the conclusions in the paper are present in the paper and/or the Supplementary Materials.

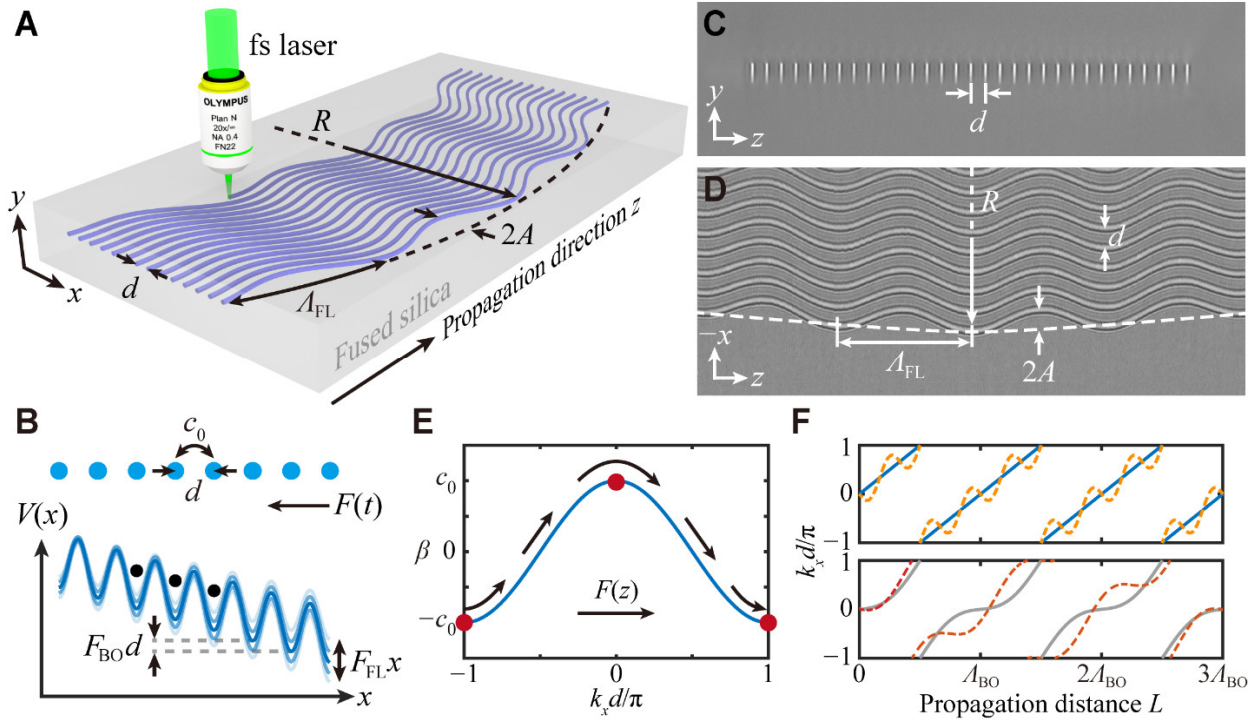

**Fig. 1. Photonic implementation and generalized acceleration theory.** (**A**) Schematic of a one-dimensional lattice composed of evanescently coupled waveguides with combined bending trajectory. (**B**) Schematic of a reduced Floquet lattice in the transformed coordinate frame. (**C**) Cross-sectional optical microscope image of the fabricated sample. (**D**) Top-view optical microscope image of the fabricated sample with a harmonic modulation. (**E**) Representation of $F(z)$ induced wave vector shift according to the generalized acceleration theory. (**F**) $z$-dependent shift of the transverse Bloch momentum for several specific cases corresponding to typical BOs ($A = 0$, blue solid line), QBOs ($\Lambda_{BO} = 3\Lambda_{FL}$, orange dashed line), SBO-like oscillations ($3\Lambda_{BO} = 4\Lambda_{FL}$, red dashed line), and spreading ($\Lambda_{BO} = \Lambda_{FL}$, gray solid line).

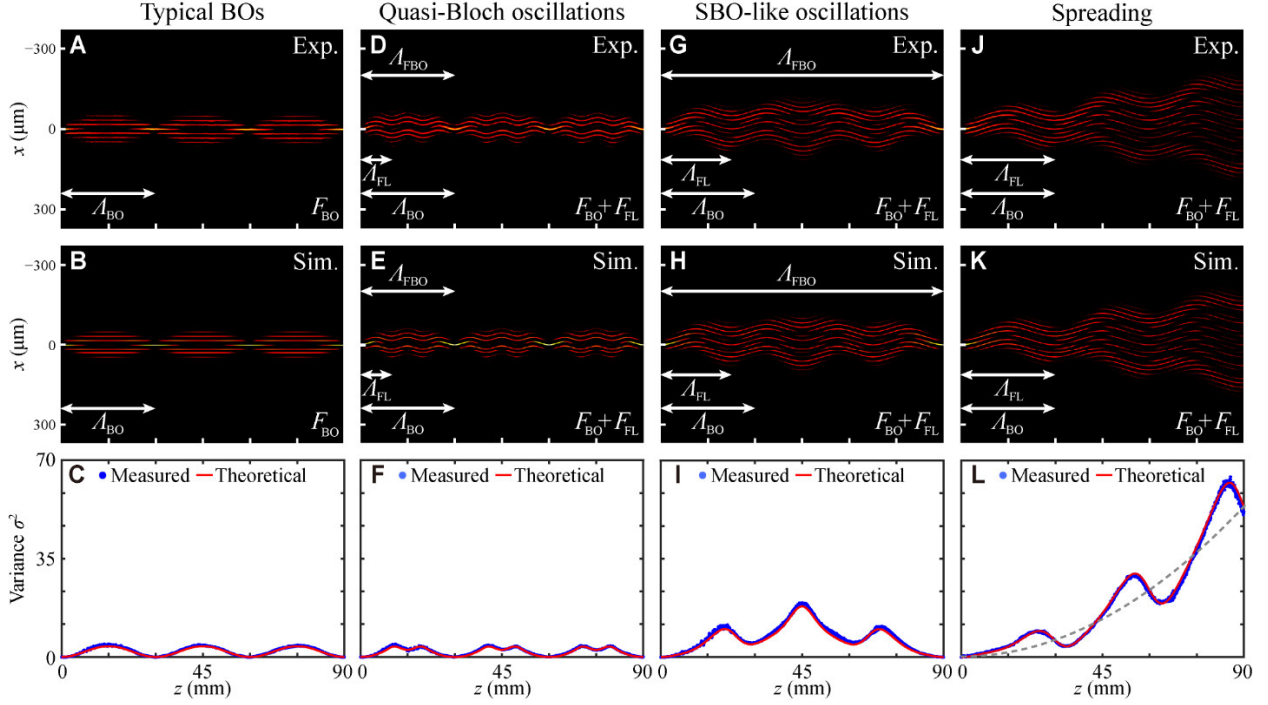

**Fig. 2. Experimental visualization, simulation, and variance of the breathing modes for single-site excitation.** (**top**) Fluorescence microscopy images of the wave evolution in curved waveguide arrays with a fixed circular bend radius $R = 110.8$ cm (corresponding to $\Lambda_{BO} = 30$ mm). (A) $A = 0$, corresponding to typical BOs; (D) $A = 6$ μm and $\Lambda_{FL} = 10$ mm, corresponding to QBO; (G) $A = 13.5$ μm and $\Lambda_{FL} = 22.5$ mm, corresponding to SBO-like oscillations; (J) $A = 18$ μm and $\Lambda_{FL} = 30$ mm, corresponding to spreading. (**middle**) Simulated the wave evolution corresponding to those in (top). (**bottom**) Corresponding variances $\sigma^2$ of the measured (top) and simulated (middle) light evolution as a function of the propagation distance $z$.

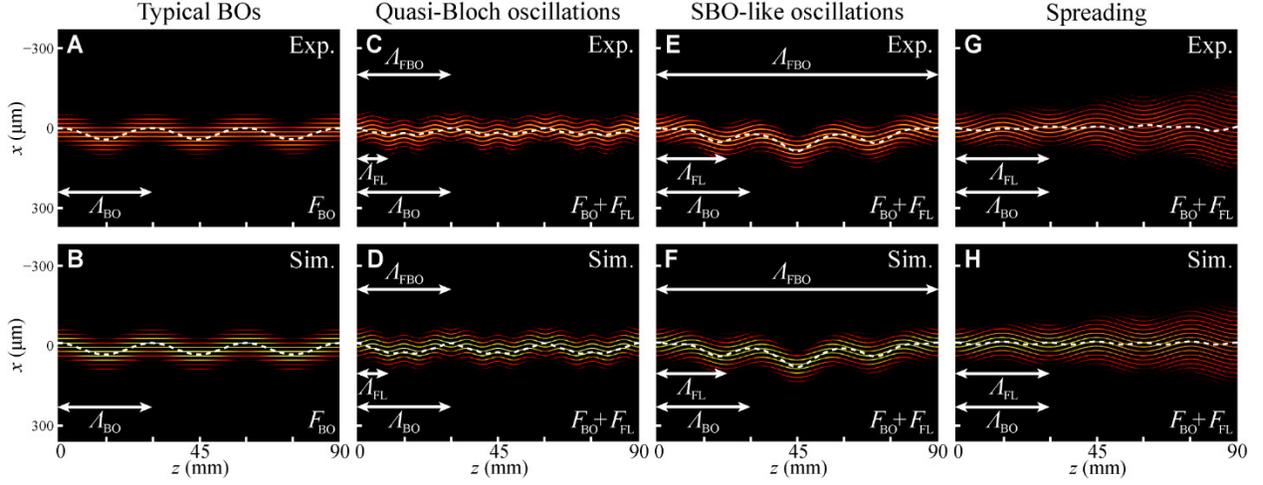

**Fig. 3. Experimental visualization and simulation of the oscillating modes for broad-beam excitation.** (**top**) Fluorescence microscopy images of the wave evolution in curved waveguides arrays with a fixed circular bend radius $R = 110.8$ cm (corresponding to $\Lambda_{BO} = 30$ mm). (A) $A = 0$, corresponding to typical BOs; (C) $A = 6$ μm and $\Lambda_{FL} = 10$ mm, corresponding to QBOs; (E) $A = 13.5$ μm and $\Lambda_{FL} = 22.5$ mm, corresponding to SBO-like oscillations; (G) $A = 18$ μm and $\Lambda_{FL} = 30$ mm, corresponding to spreading. (**bottom**) Simulated wave evolution corresponding to those in (top). The trajectories of the beam $x(z)$ extracted from the measured (top) and simulated (bottom) light evolution are marked as white dashed lines.

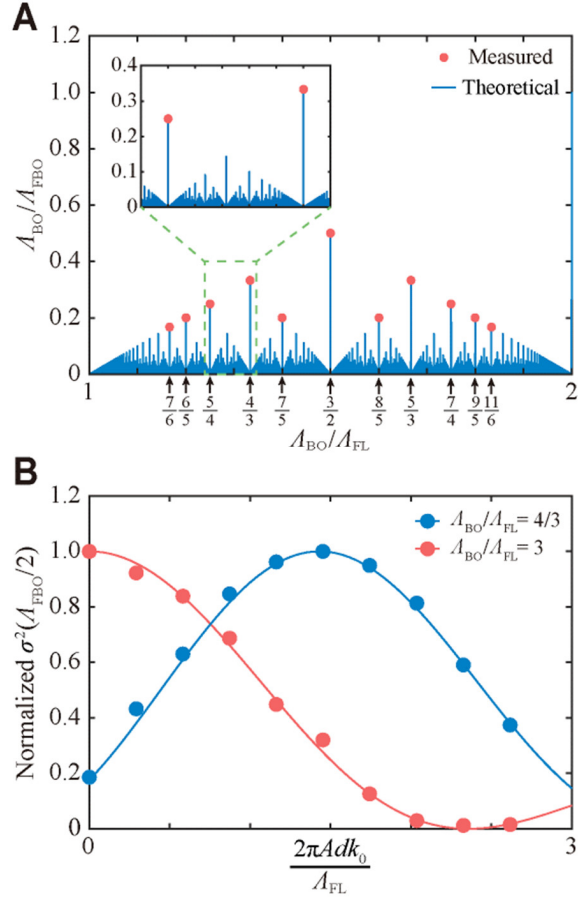

**Fig. 4. Fractal spectrum and fractional Floquet tunneling of FBOs.** (**A**) Theoretical (blue stems) and measured (red dots) ratio $\Lambda_{BO}/\Lambda_{FBO}$ as a function of the ratio $\Lambda_{BO}/\Lambda_{FL}$. The inset is a close-up spectrum at a finer scale, which show the property of self-similarity of this spectrum. (**B**) Normalized theoretical (lines) and measured (dots) FBO amplitude $\sigma^2(\Lambda_{FBO}/2)$ as a function of the ratio $A/\Lambda_{FL}$.

18